\documentclass[longauth]{aa} 

\usepackage{graphicx}
\usepackage{txfonts}
\usepackage[]{natbib}
\bibpunct{(}{)}{;}{a}{}{,}

\begin{document}
  \title{The 72-Hour WEBT Microvariability Observation of Blazar S5 0716$+$714 in 2009}

\author{G. Bhatta\inst{1}
 \and J. R. Webb\inst{1}
 \and H. Hollingsworth\inst{1}
\and S. Dhalla\inst{1}
\and  A. Khanuja\inst{1}
\and  R. Bachev\inst{2}
\and  D. A. Blinov\inst{4}
\and M. B\"ottcher\inst{6}
\and  O. J. A. Bravo Calle\inst{4}
\and  P. Calcidese\inst{6}
\and  D. Capezzali\inst{7}
\and  D. Carosati\inst{7,8,9}
\and R. Chigladze\inst{10}
\and  A. Collins\inst{11}
\and J. M. Coloma\inst{12}
\and  Y. Efimov\inst{13}
\and  A. C. Gupta\inst{14}
\and S-M. Hu\inst{15}
\and O. Kurtanidze\inst{10,16,17}
\and A. Lamerato\inst{5}
 \and  V. M. Larionov\inst{4,15,18}
\and  C. -U. Lee\inst{19}
\and E. Lindfors\inst{20}
\and B. Murphy\inst{21}
\and  K. Nilsson\inst{20}
\and J. M. Ohlert\inst{22}
\and A. Oksanen\inst{23}
\and  P. P\"a\"akk\"onen\inst{24}
\and J.T. Pollock\inst{25}
\and B. Rani\inst{3}
\and R. Reinthal\inst{20}
\and D. Rodriguez\inst{26}
\and  J.A. Ros\inst{27}
\and P. Roustazadeh\inst{5}
\and  R. Sagar\inst{14}
\and A. Sanchez\inst{28}
\and P. Shastri\inst{18}
\and A. Sillanp\"a\"a\inst{20}
\and  A. Strigachev\inst{2}
\and  L. Takalo\inst{20}
\and  S. Vennes\inst{29}
\and M. Villata\inst{12}
\and  C. Villforth\inst{20}
\and J. Wu\inst{30}
\and X. Zhou\inst{31} }

 \institute{Florida International University\\ 
              \email{gopal.bhatta@fiu.edu}
 \and   
Institute of Astronomy, Bulgarian Academy of Sciences, 72, Tsarigradsko Shosse Blvd., 1784 Sofia Bulgaria\\   
\and
 Max-Plank-Institut fur Radioastromomie Auf dem Huegel 69 53121 Bonn, Germany\\ 
\and 
 Astronomical Institute, St. Petersburg State University, Universitetsky pr. 28, Petrodvoretz, 198504 St. Petersburg, Russia.\\  
\and 
 Astrophysical Institute, Department of Physics and Astronomy, Ohio University, Athens, OH 45701, USA\\  
\and  
 Observatorio Astronomico della Regione Autonoma Valle d'Astoa, Italy\\   
\and  
 Armenzano Astronomical Observatory, Italy\\   
\and  
EPT Observatories, Tijarafe, La Palm, Spain\\ 
\and 
 INAF, TNG Fundacion Galileo Galilei, La Palm, Spain\\  
\and  
 Abastumani Observatory, Mt. Kanobili, 0301 Abatsumani, Georgia\\   
\and  
 Cork, Ireland\\  
\and  
 INAF, Osservatorio di Torino, Italy\\    
 \and 
 Crimean Astrophysical Observatory, Crimea, Ukraine\\   
\and  
 Aryabhatta Research Institute of Observational Sciences (ARIES), Manora Peak, Nainital, 263 129, India\\   
\and 
 Pulkovo Observatory, St.-Petersburg Russia\\  
\and  
 Engelhardt Astronomical Observatory, Kazan Federal University, Tatarstan, Russia\\ 
\and  
Landessternwarte Heidelberg-K\"onigstuhl, Germany\\ 
\and 
 Issac Newton Institute of Chile, St.- Petersburg Branch\\  
\and 
 Korea Astronomy and Space Science Institute, Daejeon 305-348, Republic of Korea\\  
\and 
 Tuorla Observatory, Department of Physics and Astronomy, University of Turku, V\"ais\"al\"antie 20, 21500 Piikki\"o, Finland\\   
\and
   Butler University, Indianapolis Indiana, USA\\  
\and 
 Astronomie Stiftung Tebur, Fichtenstrasse 7, 65468 Trebur, Germany\\   
\and
  Hankasalmi Observatory, Finland\\   
\and 
 Jakokoski Observatory, Finland\\   
\and  
 Dark 
Sky Observatory, Applachian State University, USA\\   
\and  
 Guadarrama Observatory, Spain\\   
\and  
Agrupacio Astronomica de Sabadell, Spain\\   
\and  
MPC-442 Gualba Observatory\\   
\and  
 Florida Institute of Technology, Melbourne Florida, USA\\   
\and 
 National Astronomical Observatories, CAS, China\\ 
\and 
  Department of Astronomy , Beijing Normal University, China\\   
      }

   \date{Received xxxx; accepted 2013}

 
  \abstract
    {The international whole earth blazar telescope (WEBT) consortium planned and carried out three days of intensive micro-variability observations of S5 0716$+$714 from February 22, 2009 to February 25, 2009.  This object was chosen due to its bright apparent magnitude range, its high declination, and its very large duty cycle for micro-variations.}
   {We report here on the long continuous optical micro-variability light curve of 0716+714 obtained during the multi-site observing campaign during which the Blazar showed almost constant variability over a 0.5 magnitude range.  The resulting light curve is presented here for the first time.  Observations from participating observatories were corrected for instrumental differences and combined to construct the overall smoothed light curve. }
   {Thirty-six observatories in sixteen countries participated in this continuous monitoring program and twenty of them submitted data for compilation into a continuous light curve. The light curve was analyzed using several techniques including Fourier transform, Wavelet  and noise analysis techniques.  Those results led us to model the light curve by attributing the variations to a series of synchrotron pulses. }
   {We have interpreted the observed microvariations in this extended light curve in terms of a new model consisting of individual stochastic pulses due to cells in a turbulent jet which are energized by a passing shock and cool by means of synchrotron emission. We obtained an excellent fit to the 72-hour light curve with the synchrotron pulse model. }
   {}


\maketitle
%

\section{Introduction}

 Blazar S5 0716+714  (DA 237, HB89) is a  BL Lac with a redshift of $z=0.3\pm 0.08$ \citep{Nil08}. It is located at high declination (+71) and is usually bright which makes it an ideal candidate for micro-variability observations.  Correlated intraday variations in optical and radio frequencies have been reported for this source \citep{quir91}.  It is also a gamma-ray emitting object associated with an FR II radio source \citep{wag96,von95,chen08,vill08}.  Observations over a period of ten years indicate it experiences nearly continuous micro-variability activity \citep{hum08,webb10}, with duty cycle of about $95.3\%$.  

Attempts to find periodicities or significant correlations have been done with extensive variability data. \citet{nesci05} used long term light curves from 1953 to 2005 and reported finding evidence of a long-term variability consistent with a precessing jet. \citet{rait03} reported a ~3.3 year period in their data.  Examination for periodicity in the short-term variability time frames yielded high significance in peaks of ~25 and ~73 minutes in twenty-two well sampled micro-variability curves \citep{gup09}.  

Among the numerous microvariability studies of this object in the literature, \cite{vill00} performed a very densely sampled WEBT observation and obtained 635 data points over 72 hours.  They found that the steepest rise and the steepest decline were both roughly 0.002 magnitudes per minute over several hours with no apparent difference between the rise and decline rates. \cite{wu05} performed nearly simultaneous multi-frequency monitoring over a span of seven days and time series analysis of their microvariability observations yielded no repeatable periodicities common in the light curves.  \cite{wu07} used an objective prism set up to obtain truly simultaneous multicolor observations over several nights. \cite{mont06} monitored the source for microvariability between 1996 and 2003 finding the most rapid variations were on the order of 0.1 magnitudes per hour over a period of two hours.  They also found no difference between the rise and decline rates.  Fourier Transform and Wavelet analysis were performed on twenty-three independent microvariability curves obtained at the SARA Observatory between 1998 and 2005, but analysis failed to yield any of the previously reported periods in the data \citep{dha10}.  In an attempt to determine if the variations in the SARA data more accurately represented by noise, \citet{dha10} applied the statistical methods of \citet{vau03} which involved comparing the RMS variations of the flux over sub-samples of the data.  The results proved inconclusive and simulations showed that even if the variations were due to noise, more data was necessary to effectively determine the noise content of the microvariations.  

Since both period and noise analysis techniques suggested that the individual microvariability curves were not of sufficient length to determine the nature of the microvariations, we organized an international campaign through the Whole Earth Blazar Telescope (WEBT) to observe S5 0716+714 over a three day period. The WEBT is a group of observatories that collaborate on blazar projects ranging from satellite back-up to campaigns such as this.  More information about the WEBT group can be found at: http://www.oato.inaf.it/blazars/webt/. We requested WEBT observers around the world observe S50716+714 during this period using standard photometric techniques, a common set of comparison stars, and in a common filter.  We report here on the data acquired during this observation (section 2), the resulting time series analysis (section 3), and we interprete the results in section 4.


\section{Observations}
\begin{table*}[ht]
\caption{Observatories contributing observations to the WEBT campaign.}             
\label{table:1}      
\centering                          
\begin{tabular}{r l l l r l c}        
\hline\hline                 
Zone & Code & Country & Observatory & Longitude & Telescope &  Filters\\    
\hline                        
   1 & AAS & Spain & Agrupacion Obs. & 0.73 & 0.5 m & R \\     
   1 & AVO & Italy & Aosta Valley Obs. & 7.36 & 0.81-m & RI \\
   1 & MAO & Germany	 & Michael Adrian Obs. & 8.41 & 1.2-m & BVRI \\
   2 &  AAO & Italy & Armenzano Obs. & 12.69 &36-cm &R\\
   3 & TUR & Finland & Tuorla Observatory & 22.17 & 35-cm, 1-m & R \\
   3 & BEL & Bulgaria & Belogradchik & 22.60 & 60-cm & BVRI \\
   3 & HANK &  Finland	& Hankasalmi & 26.50 & 40-cm (RC) &	BVRI \\
   3 & STPET & Russia & St. Petersburg & 29.82 & 40-cm & R \\
   3 & JAKO & Finland & Jakokoski Obser & 30.00 & 20-inch & I R \\	
   4 & CRIM & Crimea & Crimean AP Obser. & 	30.20 & 2.6, 1.25m & R \\	
   5 & ABAS & Georgia, FSU & Abastumani Obs. & 42.80 & 0.7-m & R\\
   6 & ARIES & India & ARIES & 71.68 & 1.04 –m & R\\
   7 & BAO & China & BAO China, Xinglong & 114.00 & 1.0-m & R\\
   7 & WHO & China & Weihai China & 122.00 & 1-m & BVRI\\
   8 & LOAO & USA & Mt. Lemmon & 249.00 & 1.0-m & R\\
 11 & MDM & USA & MDM Kitt Peak & 249.00 & MDM 1.3m & R \\	
 11 & SARA & USA & SARA/Kitt Peak & 249.00 & 1.-m & R\\
 12 & BUO & USA & Butler & 273.55 &  0.96-m & R\\
 12 & DSO  & USA & Dark Sky, North Carolina &  278.58 &24-inch & R\\
 13 & BLK & Ireland & Cork & 352.00 & 40-cm & RGB\\
\hline                                   
\end{tabular}
\end{table*}
We selected a February observation date when S5 0716+714 would be readily accessible the entire night from most Northern Hemisphere sites.  The response to the call for observers through WEBT was excellent and plans were made to carry out the observations.  The primary observing band was chosen to be R since most observatories own and regularly use that filter for microvariability observations.  We divided up the observatories into longitude regions around the globe to help coordinate the observations.  It was also decided that if there were multiple telescopes in each longitude range, then at least one telescope would be assigned a different filter so some simultaneous color information could be acquired as the observation progressed. However, the continuity of the R light curve was the highest priority in this campaign.  In addition to the common filter system, we also selected four comparison stars from the sequence of \citet{vill98}.  We chose to use stars 3, 4, and 5 as comparison stars and measuring the magnitude of star 6 as a “check” star in addition to the object.  Stars 1 and 2 frequently are overexposed due to their brightness and cause problems when trying to get accurate photometry, so they were left out of the comparison sequence.  Table 1 lists the longitude zone in column 1, the code assigned to each observatory in column 2, the location of the observatory and observatory name and longitude in columns 3,4 and 5 respectively. The telescope aperture and the filters used are in columns 6 and 7.

 \begin{figure*}[ht] 
  \centering
 \includegraphics[width=16cm,height=11cm]{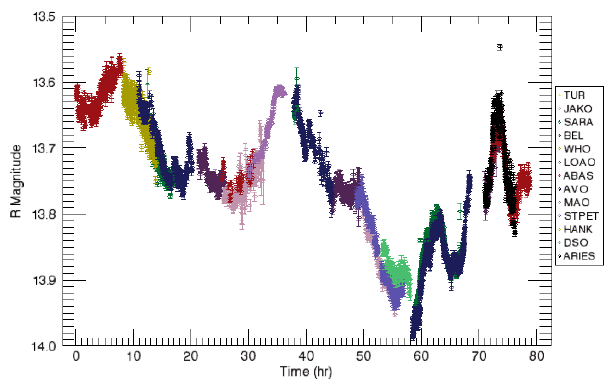}
      \caption{The raw light curve of S5 0716+714 obtained by the compilation of some of the high quality data by major contributors.  The light curves contributed from each observatory is plotted together with different symbols indentifying the observatory according to the codes given in Table 1.
              }
 \label{rawlightcurve}
  \end{figure*}
%
Figure 1 shows the raw observations plotted together.  The observations covered the time period between JD 2454886.1 (2/23/2009) and JD 2454889.5 (2/26/2009).  There was nearly continuous coverage between JD 2454887.3 and 2454888, with overlap from several observatories during many time intervals.  The code given in Table 1 is the key to the observatories responsible for each data segment on the plot.  The data were reduced at each individual observatory and sent as magnitude files for final compilation and analysis. Overlaying each contributed light curve revealed some offsets due to instrumental/filter differences, but most of the data was of sufficient quality and had sufficient overlap so that minor zero-point adjustments could be made to obtain a consistent continuous light curve.  In all cases, the data with the lowest noise and longest overlap with other data sets were used to determine the offsets for the other light curves.  Exposure times for individual images ranged from 30 seconds to 120 seconds depending on the observatory and telescope.

In order to prepare the light curve for time series analysis, we implemented a smoothing algorithm. The data were smoothed based on the assumption that given a time scale as short as two minutes, a data point cannot be too different from the average of the previous and the following point using the algorithm that if any data point is $x_i > (x_{i-1} + x_{i+1})/2 +0.005$ then, $x_i = (x_{i-1} + x_{i+1})/2$. The resulting smoothed data set consisted of 2613 high quality data points. The smoothing algorithm only affected timescales on the order of a few minutes, much shorter than any possible periods we could find from a time series analysis. In order to ensure there was no unexpected bias introduced by the smoothing, we also preformed all of the time series analysis reported below on the unsmoothed data the results were identical to those performed on the smoothed data in the frequencies of interest.  The magnitudes were converted to flux using standard flux conversions for the R filter as given by \cite{johnson88} using a redshift of 0.30 and Galactic absorption of 0.031 magnitudes.  Figure 2 shows the complete smoothed flux curve.  Smoothing kept the major trends of light curve intact, while cutting out some of the high frequency noise in the data.  The total length of the light curve was 78.88 hours. 

We analyzed individual segments of the flux curve to determine the maximum climb and decline rates.  Twelve individual rapid excursions were noted in the flux curve and we fit a line to each of those segments to determine the maximum climb rates and decline rates, concentrating on segments which had a large number of data points.  The fastest rate was an increase of 0.089 magnitudes per hour over a range of 0.15 magnitudes.  The correlation coefficient for that fit was $r^{2} = 0.997$ and it contained 99 data points. Table 2 lists the slopes and correlation coefficients for each of the segments we examined.  Column 1 is the segment index, column 2 the start and finish time of the segment in hours, and column 3 the slope in magnitudes per hour.  We calculated the correlation coefficient for each fit and listed them in column 4 along with the number of points in column 5.  Column 6 gives the probability that a random sample would show such a large correlation coefficient.  The final column denotes whether the slope is a rise or a decline in magnitude.     
\begin{table*}
\caption{Maximum slopes of the various sections from the light curve.}             
\label{table:2}      
\centering                          
\begin{tabular}{l c l l c cl}        
\hline\hline                 
Seg&Times & Slope &  $r^{2}$  & Npts & Prob. & Rise/Decline \\    
 & (hr) & (mag/hr) &  &     &   & \\
\hline                        
  1&   3.38-7.01 & 0.019 & 0.918 & 123 & $5x10^{-7}$ & Rise\\  
2&13.03-16.54 & 0.027 & 0.946 & 155 & $5x10^{-7}$ & Decline\\ 
3&18.69-19.84& 0.064 & 0.967 & 40 & $1x10^{-6}$ & Rise\\ 
4&29.91-35.76 & 0.029 & 0.968 & 120 & $1x10^{-6}$ &  Rise\\ 
5&38.39-39.91 & 0.050 & 0.945 & 77 & $5x10^{-7}$ & Decline\\ 
6&49.70-55.30 & 0.022 & 0.956 & 177 &  $5x10^{-7}$ & Decline\\ 
7&58.52-63.10 & 0.037 & 0.878 & 244 & $5x10^{-7}$ & Rise\\ 
8&63.48-65.20 & 0.033 & 0.904 & 115 & $5x10^{-7}$ & Decline\\
9&66.74-68.46 & 0.089 & 0.977 & 99 & $5x10^{-7}$ & Rise\\
10&72.14-73.60& 0.035 & 0.739 & 29 & $1x10^{-6}$ &   Rise \\
11&74.07-75.63 & 0.076 & 0.977 & 43 & $1x10^{-6}$ & Decline\\
12&75.66-78.17 & 0.026 & 0.873 & 73 & $5x10^{-7}$ &  Rise\\   
\hline                                   
\end{tabular}
\end{table*}
The average decline rate was 0.042 magnitudes/hour (standard deviation of 0.022) while the average rise rate was 0.043 with a standard deviation of 0.028.  Thus overall, the rise and decline rates are similar in this segment of light curve. Although the rates are different, the fact that the slopes for the rise and decline are the same agree with the results found by \citet{vill00,mont06}.
\section{Time Series Analysis}
\subsection{Fourier Transform Analysis}
We performed Fourier Transform analysis on the entire smoothed light curve by removing the linear trend of slope $-2.5\times10^{-5}$ mJy/hr and using a Discrete Fourier Transform algorithm \citep{dem75}. The results of this analysis are shown in Figure 3 and Table 3. The DFT results yielded some of  the  large features at periods of 40.00, 21.05 and 13.19 hours which correspond to the first, second and fourth peaks respectively in Fig. 3. Some of the periods which are well above the noise level are listed in column 4 of Table 3.  The corresponding time-scales in the rest frame in column 5 were calculated using

\begin{equation}
\Delta t_{rest}=\frac{D}{1+z}\Delta t_{obs},
\end{equation} 
where $D=1/\Gamma \left (  1-\beta cos\theta \right )$ is Doppler factor with  $\beta c$ bulk speed  and $\Gamma$ bulk Lorentz factor. $\theta$ is the orientation of the jet axis with respect to the line of sight and $z$ is the red-shift of the object. Using the values $\Gamma =17$ , $\theta=2.6^{o}$  and $z=0.3$, a Doppler factor of 21.32 was used to transform the time-scales in the rest frame \citep[see][]{celotti08}.

 The horizontal line across the bottom indicates the average power of 100 light curves generated in Interactive Data Language (IDL) assuming  Gaussian distributed noise with sigma equal to the sigma of the flux curve.  The peaks in the S5 0716+714 DFT are thus many sigmas above the noise level. We then pre-whitened the flux curve by subtracting the derived period, adjusting the amplitude and phase for best fit, and then reanalyzed the resulting light curve using the DFT.  The amplitude of peaks in the DFT did not decrease drastically as we pre-whitened the data, indicating that the periods are not indicative of a true period running through the extended data set.  Neither did we find the periods at approximately 25 and 73 minutes in this flux curve seen previously by \cite{gup09}.  Thus we could not confirm any of the previously detected periods seen in the $S5 0716+714$ microvariability curves or propose significant new periods.  We attribute the seemingly significant peaks seen in the DFT to particular features in pieces of the flux curve, not cyclical oscillations throughout the entire 72-hour observation. 
 \begin{figure}[h]  
   \centering
  \includegraphics[width=9cm]{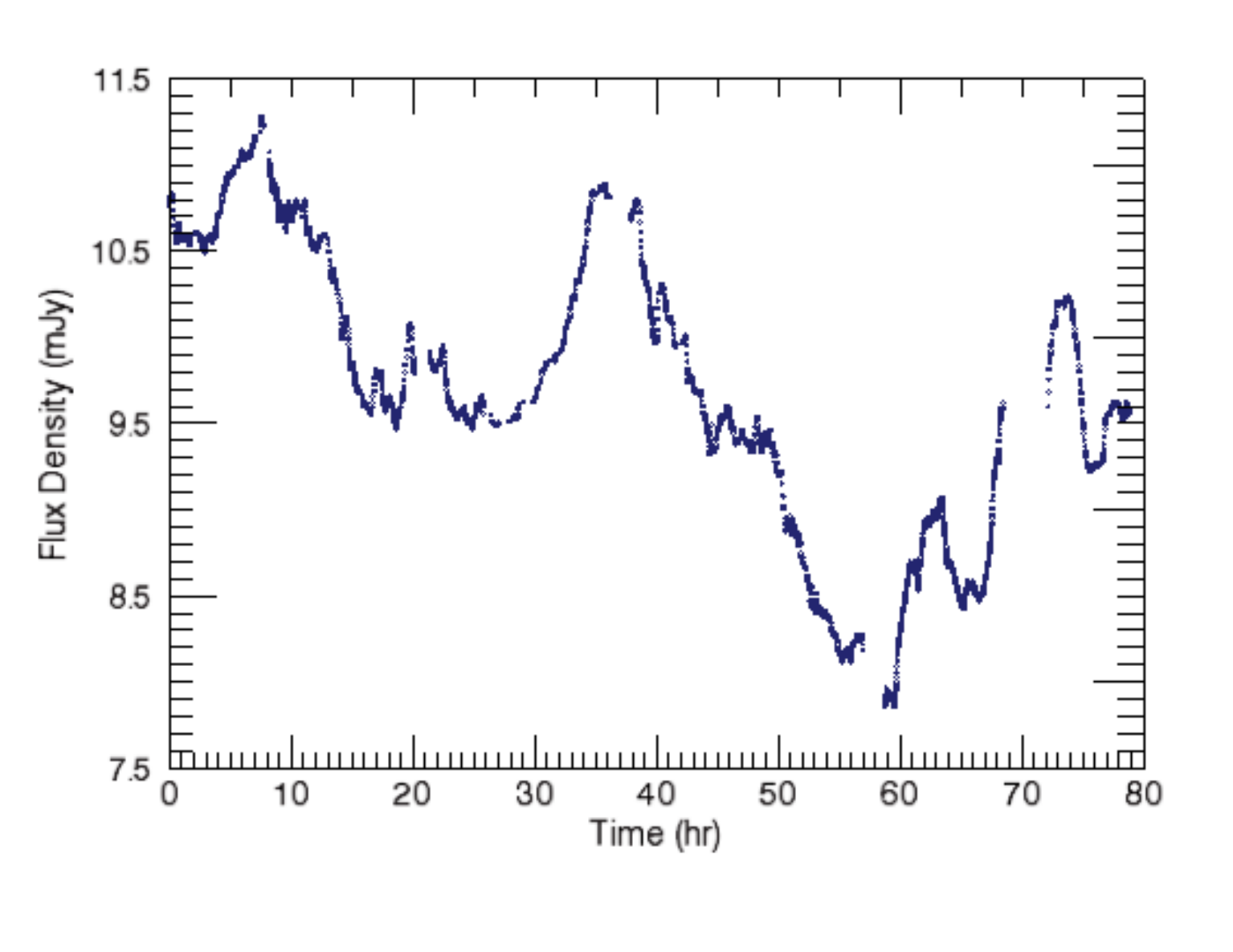}
      \caption{The flux curve of 0716+714 smoothed using the algorithm discussed in the text. }
         \label{smoothedlightcurve}
   \end{figure}
\begin{figure}[h]
\centering
\includegraphics[width=9cm]{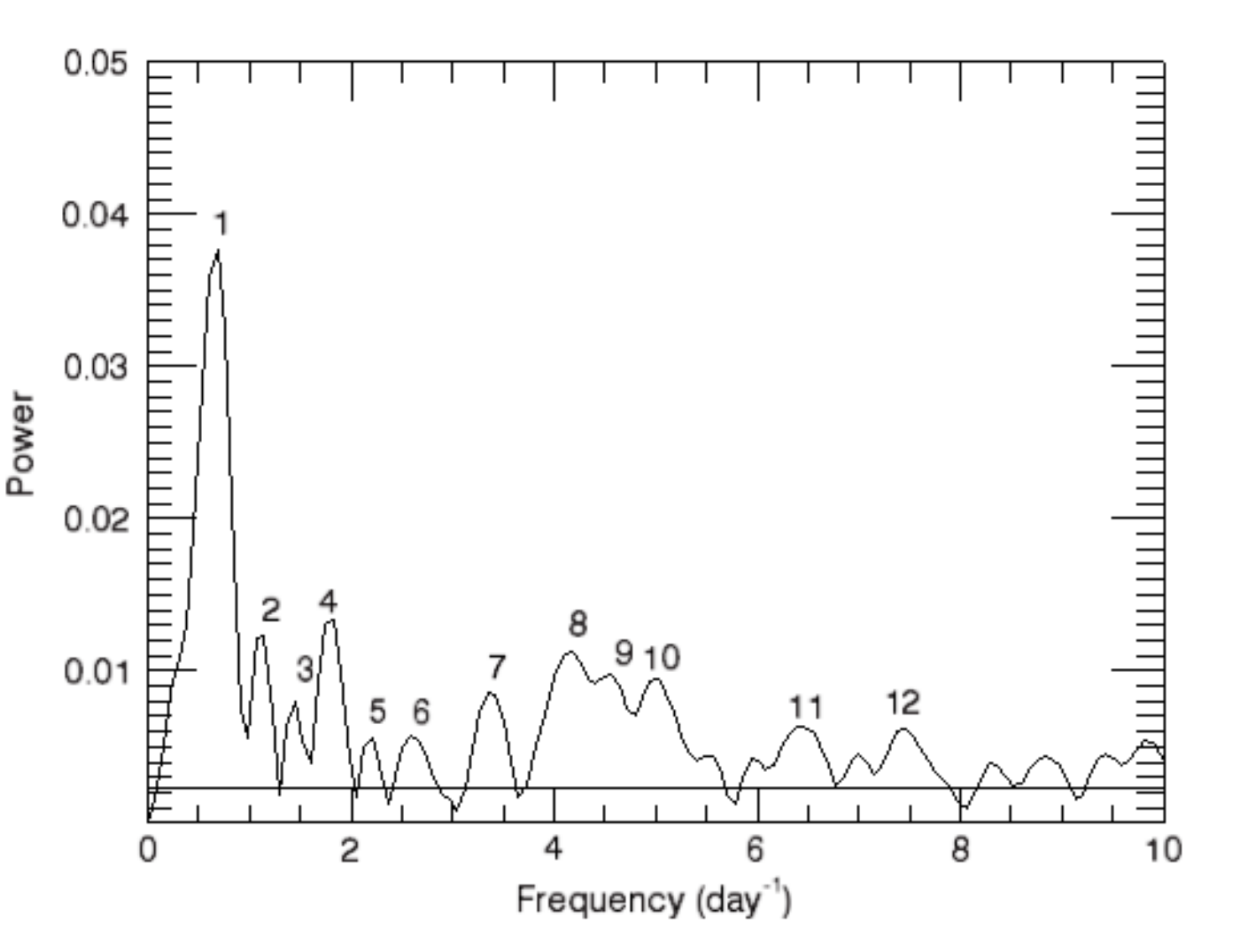}  
\caption{The power spectrum of the smoothed Light curve. The horizontal line at low power indicates the maximum power level of simulated random noise light curves. The numbers associated with the peaks in the power spectrum correspond to the frequencies and periods listed in Table 3.}
\label{powerspectrum}
\end{figure}

\begin{table*}
\caption{Periods along with their corresponding timescales in the observed  and rest frame as detected with DFT analysis. }             
\label{table:3}      
\centering                          
\begin{tabular}{r c r c c}        
\hline\hline                 
Number & Frequency & Power & Timescale in obs. frame& Timescale in rest frame \\    
   &       (/day)   &     & ( hr)  &   ( hr) \\ 
\hline                        
 1 & 0.60 & 0.0382 & 40.00 & 656.00   \\  
2 & 1.14 & 0.0122 & 21.05  & 345.22      \\ 
3 & 1.41 & 0.0083 & 17.02  & 279.13  \\ 
4 &  1.82 &  0.0126 & 13.19  & 216.32    \\ 
5 & 2.25& 0.0052 & 10.67  & 174.99  \\ 
6 & 2.60& 0.0054& 9.23  & 151.37 \\ 
7  & 3.40 & 0.0081& 7.06 & 115.78   \\
8 & 4.20 & 0.0121 & 5.71  & 93.64   \\
9 & 4.54 & 0.0092 & 5.29  & 86.76  \\
10 & 5.00  &  0.0091 & 4.80  & 78.72    \\
11 & 6.50  &  0.0063 & 3.69  & 60.52  \\
12 &  7.50 &  0.0062 & 3.20  & 52.48  \\   
\hline                                   
\end{tabular}
\end{table*}
We performed wavelet analysis \citep{torrence98} using the wavelet application in IDL to compute the wavelet transform of the data and then compared with the DFT results.  The Morlet kernel of order 6 was used to do the analysis.  Figure 4 shows the resulting wavelet transform.  
 \begin{figure*}
   \centering
 \includegraphics[width=16cm]{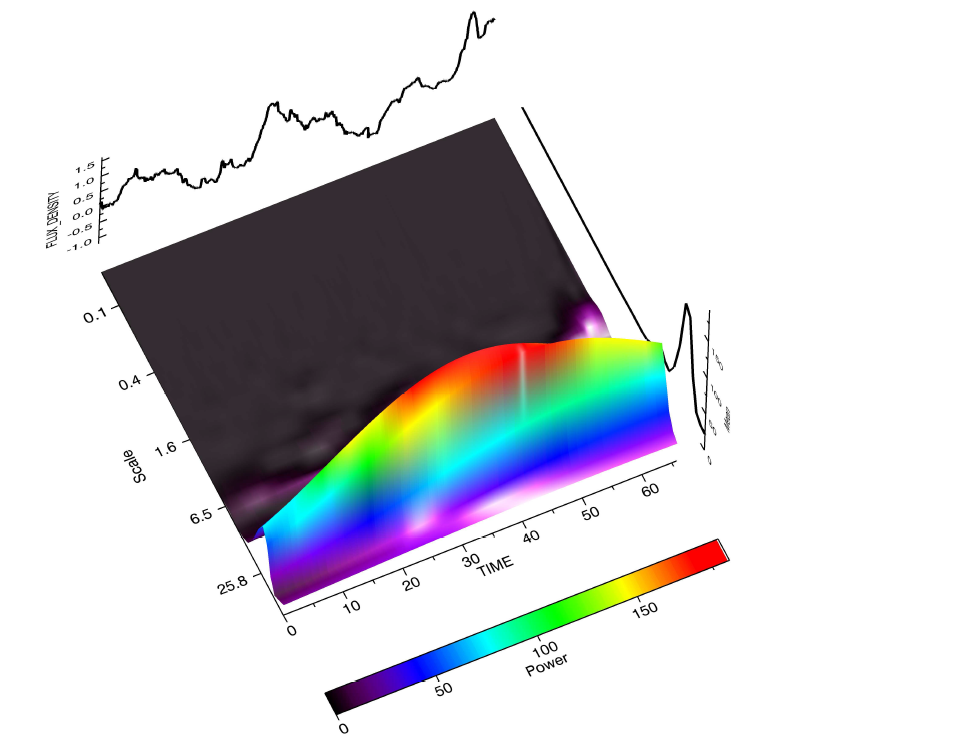}  
      \caption{Wavelet transform of the smoothed light curve linearized by removing a slope of $-2.5\times10^{-5}$ mJy/hr and a Y-intercept of flux 0.01 mJy. Morlet kernel of 6th order was used from the wavelet application in IDL. }
         \label{waveletspec}
   \end{figure*}
The peak of the wavelet transform corresponds to a sinusoidal oscillation with a period of 30.72 hours, but is clearly only significant in the center of the data run, fitting extremely poorly at the start of the light curve.  The inconclusiveness of both the DFT and wavelet analysis, along with not reproducing any of the previously reported periods in this object led us to examine the flux curve in terms of a noise model.
\subsection{Noise Analysis}
\cite{dha10} used the analysis methods of Vaughan et al. (2003) to examine twenty-one single-night microvariability light curves of S5 0716+714. This analysis consisted of dividing the flux curves up into individual bins and calculating the RMS deviation within the each bin given by:
\begin{equation}
     x_{rms} = \sqrt{\frac{1}{N}\sum_{i=1}^{N}(x_i - \bar{x})^2}.
   \end{equation}

According to \citet{vau03}, if light curves are strictly the result of Gaussian noise processes, each independent flux curve would be one realization of the underlying stochastic process.  If the process is a stationary noise process, the realizations should exhibit similar statistical properties e.g. a linear relationship between excess RMS and average flux. \citet{dha10} found that the available microvariability curves were too short to reliably recover the noise characteristics of the data and concluded that the individual microvariability curves needed to be longer to determine whether the variations are the result of a deterministic stochastic process.  The WEBT observation presented here should be long enough to identifiy the nature of the microvariations if in fact the microvariability is due to purely noise processes.  We repeated the same noise analysis reported by \cite{dha10} with the WEBT 72-hour flux curve of S5 0716+714. The data was binned into 45 minute bins and only bins with a minimum of 20 data points were used. Figure 5 shows the RMS vs. average flux plotted for the bins and it fails to show the expected linear relationship between RMS and average flux if the the variations were due to a totally Gaussian noise process. The correlation of the best fit line to the data is 0.0003.  In a further effort to deduce the noise content of the data, we re-plotted the DFT in log-log space in Figure 6.  The DFT plotted in this way is best fit to a slope of  $1/f^2$ noise, but there are several prominent features at low frequency which deviates from a simple noise distribution.   
 \begin{figure}
 \centering
\includegraphics[width=9cm]{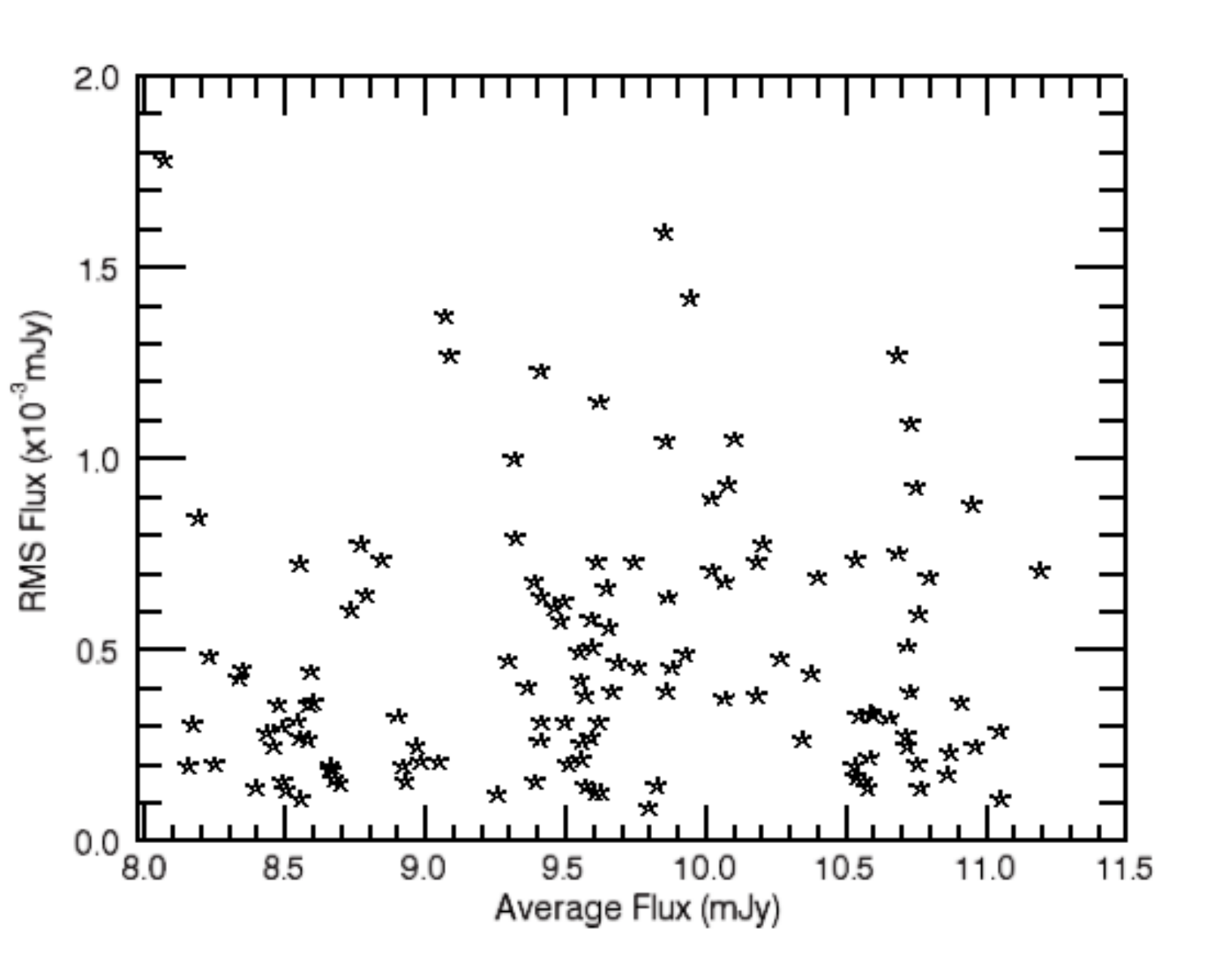}  
       \caption{The RMS variations versus average flux for the entire 3-day observation.}

         \label{rmsvariation}
   \end{figure}

The result of not verifying any of the previously detected periods and the inconclusiveness of the noise analysis has led the authors propose a new model for the interpretation of microvariability.  We propose the variations are due to stochastic pulses generated by a shock encountering a turbulent jet flow.  This model is  described in the following section then fit to the 72-hour microvariability curve.
\begin{figure}
   \centering
\includegraphics[width=9cm]{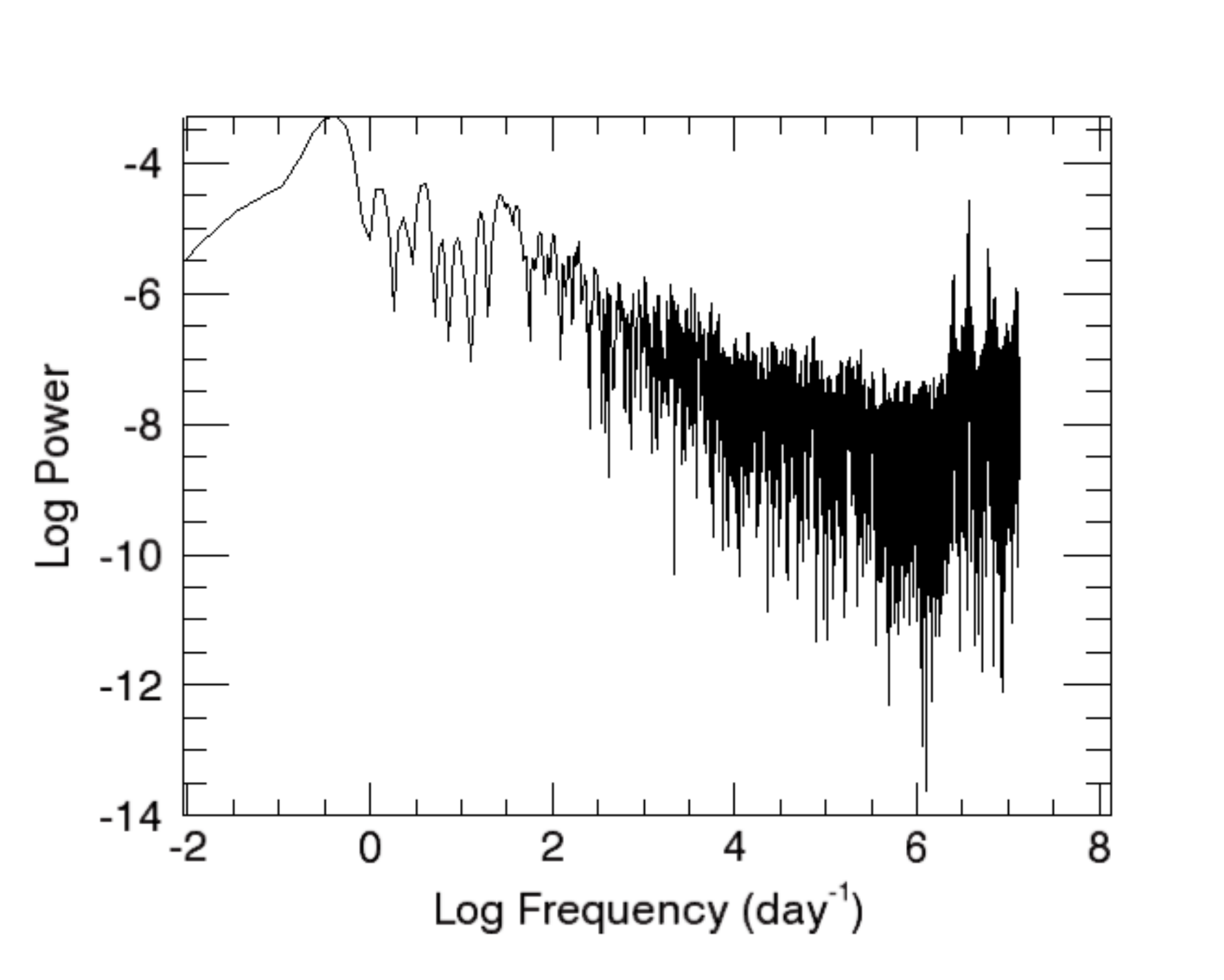}
      \caption{The log-log plot of the power spectrum of 0716+714. }
         \label{waveletspectrum}
   \end{figure}
\section{Modeling the Microvariations}

Since time series analysis of the microvariability flux curve did not show repeatable periodicities nor a strictly noise characteristics, and since close inspection of many microvariability curves show that the microvariations can be resolved into pulses or shots, we investigate a model where individual synchrotron cells are energized by a plane shock propagating down the jet and result in an increase in flux resembling a pulse. 

\subsection{Turbulence in the plasma jet}

 The most natural model for the generation of stochastic synchrotron cells is the presence of turbulence in the jet.  \citet{jones88} numerically simulated a turbulent relativistic jet using nested cells to investigate the circular de-polarization properties of the VLBI radio jets.  \citet{marsch92} investigated a turbulent model of the plasma in the radio jets discussed the interpretation of radio flickering seen in jets as a result of a shock encountering this turbulence. We are following these seminal works by assuming a turbulent jet is responsible for the microvariations and that as the shock accelerates particles in each cell, the particles cool by synchrotron emission.  The individual vortices each could have different densities, sizes and magnetic field orientations.  We further assume that as the strong shock hits each cell, rapid particle acceleration and subsequent cooling by synchrotron emission produces a pulse in the flux curve.  The convolution of these individual pulses leads to the observed  microvariability.  The formation of the turbulent cells is a stochastic process, thus producing no strict periodicities in the data.  We suppose that by de-convolving the microvariability curve into individual pulses, we can gain understanding into the underlying turbulence in the jet. To do this, we need to have a reasonable idea of the pulse profiles expected from the turbulent cells. 

\subsection{Calculation of the Synchrotron Pulse Profiles}

\cite{leht89} first investigated shot models to describe the 1/f nature of X-ray light curves.  He studied shots that had a delta function rise and exponential decline and was able to show that a random distribution of such shots yields a 1/f power spectrum.  The shot profiles studied by \cite{leht89} did not resemble the microvariability seen in 0716+714, but visual inspection of over 100 microvariability light curves between 1989 and 2009 \citep{mont06,hum08} led \citet{webb10} to propose that the microvariability curves were indeed shots, but shots with a light curve profile given by \citet{kirk98} (hereafter KRM).  KRM calculated the particle acceleration in the shock front for various magnetic field orientations and particle densities assuming a plane shock encounters a cylindrical density enhancement.  In this model, the ratio of the acceleration time $t_{acc}$ and escape time $t_{esc}$, in addition to constraints on the cooling length $L$ control the pulse shape.  The amplitude is given by the parameter $Q$ and is related to the magnetic field strength $B$ and orientation $\theta$ in addition to the enhanced electron density.  The profiles described by KRM look similar to the microvariability pulse profiles seen in the 0716+714 microvariability curves and microvariability curves of other objects \citep{bhat11}.  The general picture of this model is schematically shown in Figure 7.  We re-evaluated the KRM model for our case assuming that the electrons accelerated by the shock obey the particle distribution function given by \cite{ba92}. 
\begin{figure}
   \centering
\includegraphics[width=8cm]{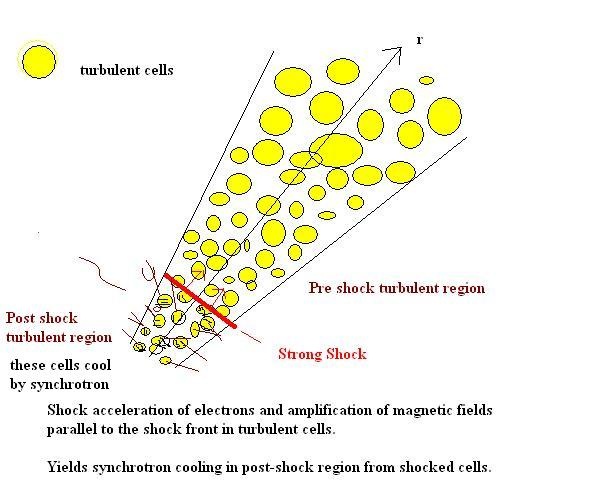}
      \caption{Example of geometry used in solving the Kirk diffusion equations. }
         \label{geometry}
   \end{figure}
\begin{equation}
\frac{\partial N}{\partial t} + \frac{\partial }{\partial \gamma }\left [ \left ( \frac{\gamma }{t_{acc}} - \beta _{s}\gamma ^{2}\right )N \right ]+\frac{N}{t_{esc}} = Q\delta \left ( \gamma -\gamma _{0} \right ) 
\end{equation}
with
\begin{equation}
\beta _{s} = \frac{4}{3}\frac{\sigma _{T}}{m_{e}c}\left ( \frac{B^{2}}{2\mu _{0}} \right )
\end{equation}  

\noindent where $N$ is the number density of the electrons in the energy space represented by Lorentz factor $\gamma$. $t_{acc}$ and $t_{esc}$ are the time of acceleration and time of escape respectively. The term $\beta _{s}\gamma ^{2}$ in Eq. 2 represents energy loss of an electron due to synchrotron emission where $\sigma _{T}$, $\mu _{0}$, $m_{e}$ and $c$ are Thomson's scattering cross-section, permeability of the free space, mass of an electron and the speed of light respectively. 

We assumed the bulk velocity of the jet material as 0.9c with the shock moving at a velocity of 0.1c relative to the jet material.  We first  determined a linear baseline in flux space for the entire microvariability curve.  This is interpreted as the background emission of the laminar flow in the jet.  The pulse code written in IDL following KRM required the solution of a transcendental equation for the cooling length and a numerical integration for the total intensity emitted by each turbulent cell.  The numerical solution of these equations were for the case where there is a constant injection rate  $Q_{0}$, before switch-on, which occurs at $t=0$.  The solution involved the variables $t_{acc}$, $ t_{esc}$, $B$, and $\gamma$ which are physical parameters that determine the pulse rise and decay time; the amplitude is determined by changing the Q value for the pulse duration.  The magnetic field B and the angle of the field relative to the line of sight $\theta$ can be different for each cell in a turbulent medium thus affecting the amplitude of the synchrotron emission and is folded into the Q parameter.  Pulses are produced by allowing $Q(t) = Q_{0}$ for $t < 0$ and $t > t_{f}$, while the strength of the pulse is $Q(t)=(1 + \eta _{f}) Q_{0}$ for $0 < t_{f}$  where $1 + \eta _{f}$ represents the factor by which the rate of injection increases during $ t_{f}$. The intensities are then calculated by $I(\nu,t) = I_{1}(\nu,\infty) + I(\nu,t_{flare})$ which is similar to Eq. 24 in KRM paper.  The value of Q during the flare determines the amplitude of the pulse, the ratio $t_{acc}/t_{esc}$ determines the shape, and the duration $t_{f}$ determines the width of the pulse.  After a number of test solutions, we determined that for  $t_{acc}/t_{esc}=0.5$  we get symmetric pulse shapes similar to what we see in the microvariability curves.  The pulse shape is shown in Figure 8. We used this as the standard pulse shape for all subsequent fits. 

\subsection{Modeling the 72-hour flux curve}
We fit thirty-five pulses to the microvariability curve by varying the width and amplitude of the standard pulse.  Figure 9 shows the light curve fitted with the convolved pulses.  The blue points are the smoothed data and the red points are the model.  The resulting parameters for the pulses used in modeling the light curve are listed in Table 4.  The first column of Table 4 shows the shot number in time order, while the second column records the center time of the pulse.  The center time is related to the relative location of the cell along the jet as the shock progresses at a velocity of 0.1c. The amplitude and width ($\tau_{flare}$) of each of pulse are given in the 3rd and 4th columns respectively, and the calculated electron density enhancement in the 5th column.  The amplitude of the pulse over the background gives the value of $Q$, which is related to either the magnetic field enhancement or the density enhancement.  The number quoted in Column 5 is calculated assuming the pulse is due to the density enhancement only.   Column 6 gives an estimate of the size for the cell in AU based on the assumed shock speed ($u_{s}=0.1c$) and the duration of the pulse. The final column is an index number determined by sorting the cell sizes in order from the smallest to the largest size. The resulting fit of the 35 convolved pulses over the baseline flux of 7.85 mJy to the 72-hour light curve had 2,507 degrees of freedom $(n_{pts} - (35\times{3})-1)$ and the fit yielded a correlation coefficient of 0.98.  Although the fit is not unique, it is representative of how well the model compares to the data.  


\begin{figure}[h]  
\centering
\includegraphics[width=9cm]{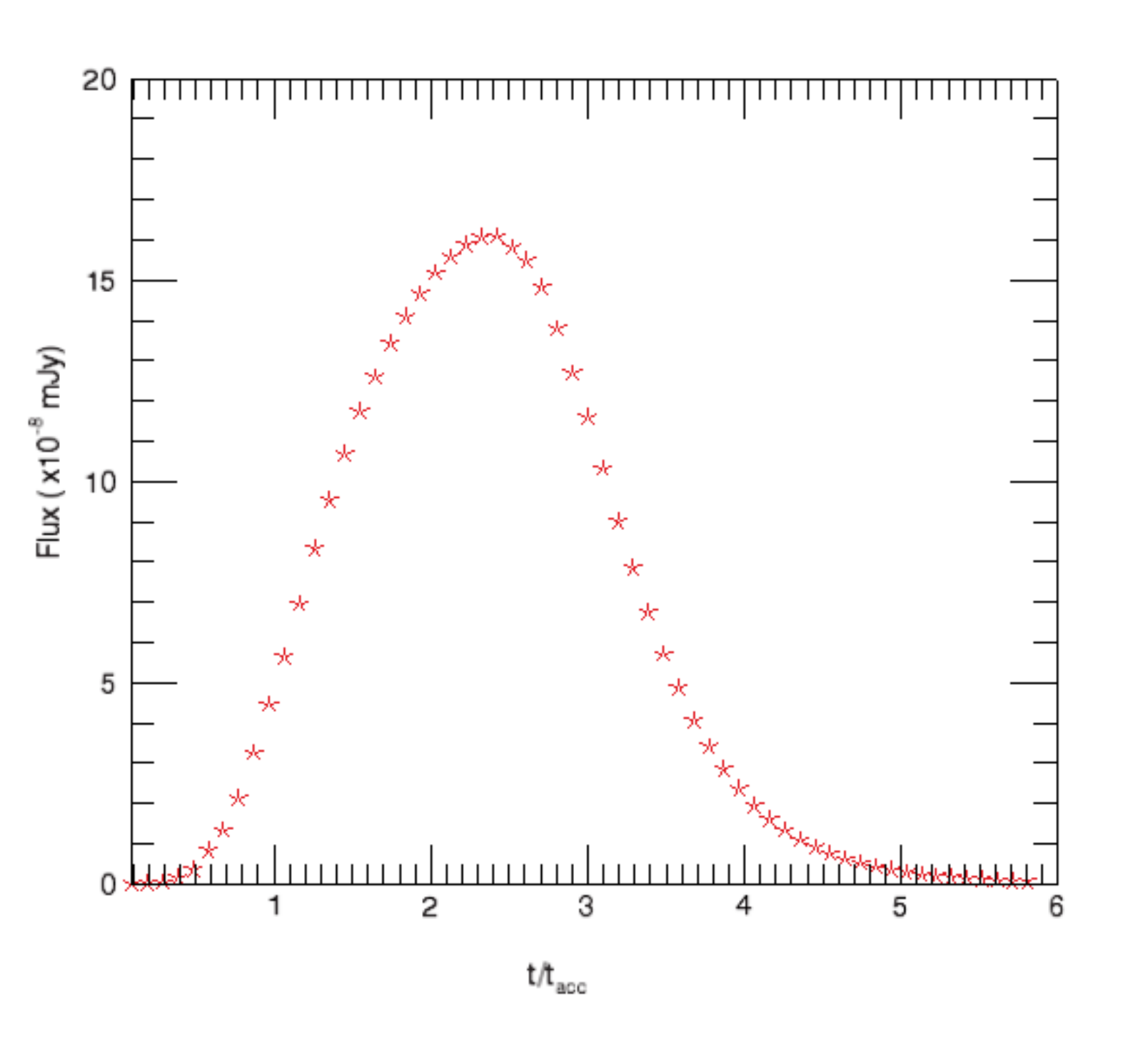}
\caption{A  single pulse emission in the local frame of reference due to the particle density and magnetic field enhancement. }
\label{Pulse Emission}
\end{figure}

\begin{table*}
\caption{The Pulse Fit Parameters}             
\label{table:4}      
\centering                          
\begin{tabular}{ r r c c c c l}        
\hline\hline                 
Pulse & Center & Amp & $\tau_{flare}$  & N & $S_{cell}$&  $Index$\\    
         &   (hrs)    &   (mJy)    &  (hrs) &  $ x10^{-5} (s^{-1} m^{-3})$ & AU & \\ 
\hline                        
        1 &  0.15 &  0.57 & 1.11&4.94&13.18&4\\
       2 &  2.10 &  2.74& 9.74&23.56&115.35&34\\
       3 &  5.70 & 1.27 & 3.34&10.94&39.54&23\\
       4 & 7.80 &  1.47 & 3.75&12.66&44.49&30\\
       5 & 8.45 &  1.24 & 2.78&10.69&32.95&18\\
       6 & 10.80 &  0.14 & 3.61&1.25&42.84&28\\
       7 & 10.90 &  2.39 & 2.92&20.56&34.60&20\\
       8 & 13.00 &  2.34 & 2.50&20.13&29.66&17\\
       9 & 14.30 &  1.29 & 1.41&11.12&16.80&7\\
      10 & 14.60 &  0.19 & 0.55&1.68&6.59&1\\
      11 & 15.50 &  1.69 & 1.67&14.55&19.77&9\\
      12 &  17.20 &  1.94 & 2.19&16.69&26.03&15\\
      13 & 18.20 &  0.72& 0.78&6.22&9.22&2\\
      14 & 19.75 &  2.04 & 2.45&17.55&29.00&16\\
      15 & 21.30 & 0.57 & 0.78&4.94&9.22&3\\
      16 & 22.30 &  1.34 & 2.03&11.55&24.05&14\\
      17 & 24.10 &  0.49 & 1.75&4.25&20.76&10\\
      18 & 25.65 &  0.34 & 1.39&2.96&16.47&5\\
      19 & 29.10 &  1.74 & 13.91&14.98&164.79&35\\
      20 & 32.40 &  0.59 & 3.47&5.11&41.19&25\\
      21 & 35.55 &  2.34 & 4.17&20.13&49.43&31\\
      22 & 38.69 &  2.44 & 2.78&20.99&32.95&19\\
      23 & 40.55 &  0.94 & 1.58&8.11&18.78&8\\
      24 & 42.60 &  2.09 & 4.45&17.98&52.73&32\\
      25 & 45.95 &  1.44 & 2.92&12.40&34.60&21\\
      26 & 48.90 &  1.64 & 3.61&14.12&42.84&29\\
      27 & 51.40 & 0.79 & 2.00&6.83&23.73&13\\
      28 & 53.75 &  0.54 & 3.20&4.68&37.90&22\\
      29 & 56.40 &  0.37 & 1.94&3.22&23.07&12\\
      30 & 60.80 &  0.64 & 1.39&5.54&16.47&6\\
      31 & 63.10 &  1.24 & 3.48&10.69&41.19&26\\
      32 & 65.85 &  0.64 & 1.86&5.54&22.08&11\\
      33 & 68.80 &  1.84 & 3.42&15.84&40.53&24\\
      34 & 73.40 &  2.39 & 5.28&20.56&62.62&33\\
      35 & 77.80 & 1.64 & 3.56&14.12&42.18&27\\
      
\hline                                   
\end{tabular}
\end{table*}

\begin{figure*}  
\centering
\includegraphics[width=18cm,height=15cm]{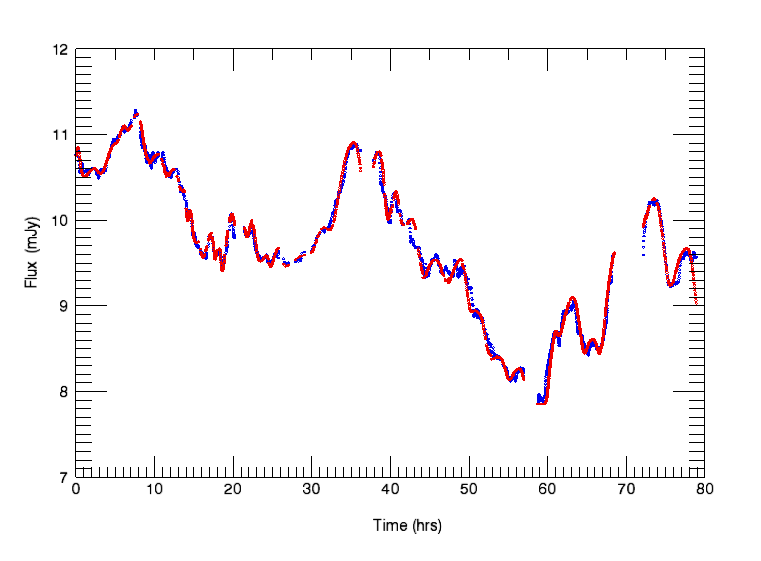}
\caption{The light curve fitted with the convolutions of the synchrotron pulses of various amplitudes and widths presented in table 4 - the curve in blue is the data and the one in red is the fit. The correlation coefficient between them is 0.98. }
\label{fitted light curve}
\end{figure*}

\begin{figure}  
\centering

\includegraphics[width=9cm]{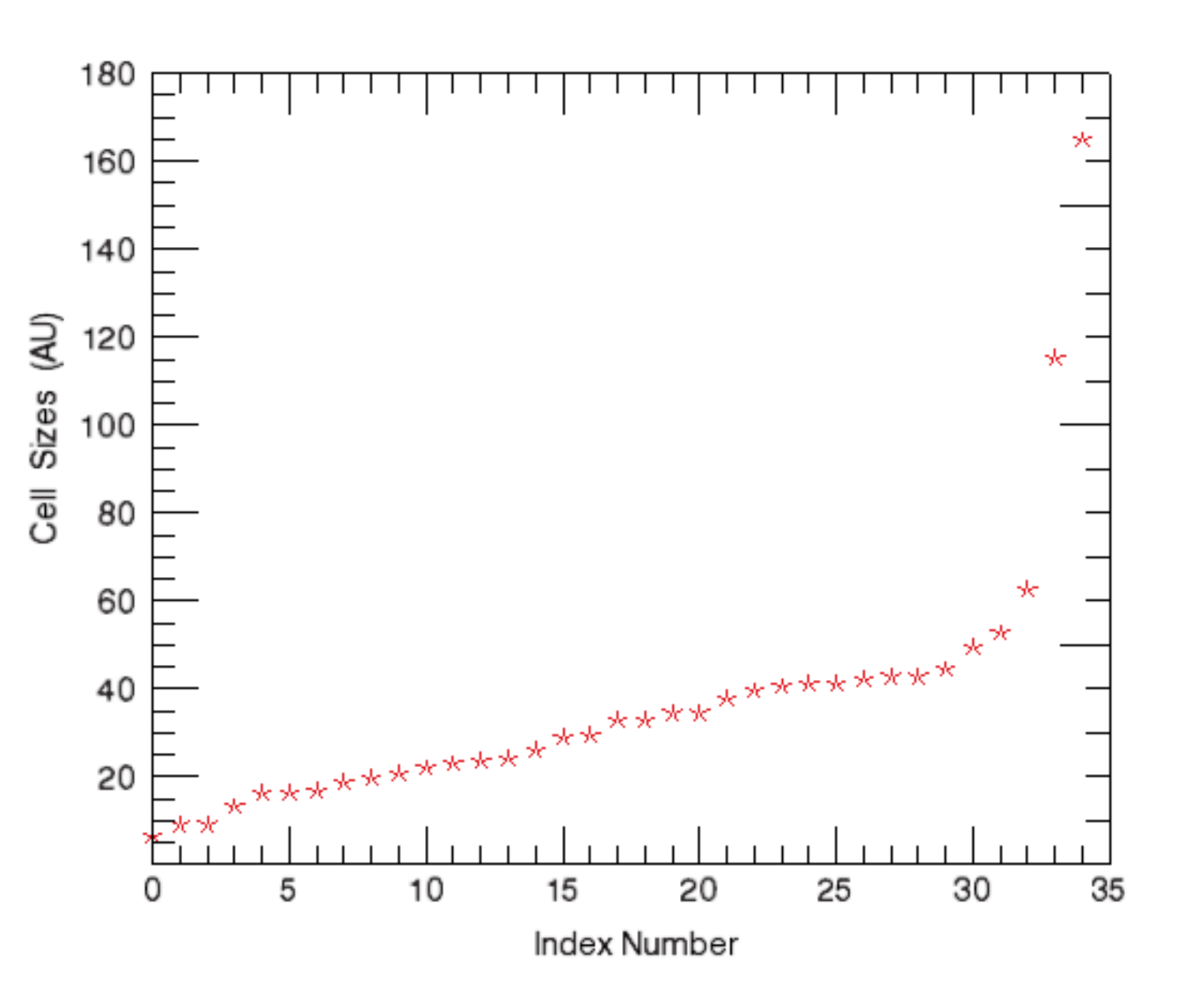}
\label{cell_dist}
\caption{The distribution of the cell sizes in AU constructed using the data from the table 4. The figure shows that the cell sizes form a continuum up to the size about 60 AU.}
\end{figure}

We can estimate the turbulent parameters from the observations by first noting that the very large duty cycle for microvariations in this object suggests the Reynolds number is well above the critical value and the plasma flow is normally turbulent.  Since turbulence is a stochastic process, each micro-variability curve is a realization of that stochastic process.  There is a large range of length/time scales for the turbulent vortices.  The smallest vortex timescale is normally associated with the Kolmogorov scale (where most of the dissipation takes place in non-relativistic plasma) and the largest length scales are associated with either the size of the plasma jet or the correlation length within the plasma.  Relativistic simulations show that turbulent relativistic extragalactic jets show a similar relationship between vortex length scale and energy \citep{zrake12} as that seen in non-relativistic plasmas. Figure 10 shows the cell size in AU plotted against index number.  The largest cell sizes are $\sim$160 AU which could either correspond to the correlation length scale or the physical width of the jet, while the smallest cell sizes seen are around 6 AU, could correspond to the Kolmogorov scale length of the turbulent plasma.  The distribution of cell sizes is continuous which is what we would expect from a turbulent jet.  Further analyses of the turbulent properties of the jet are underway by looking at other microvariability curves of this object in terms of this model.  Fitting pulses to other microvariability curves might give us a much better indication of the turbulent nature of the plasma in these sources since each microvariability curve is a realization of the turbulent plasma in the jet.

\section{Conclusions}

The WEBT provided an excellent means to use longitude distributed ground-based telescopes to continuously monitor a very active blazar with minute time resolution.  As a result of that program, we compiled a 72-hour nearly continuous light curve of $ 0716+714$.  This light curve enabled us to do detailed time series analysis on the data and to confirm the maximum rise/decline timescales seen by other observers in the microvariability curves of this object.  Based on these observations, we feel we may have resolved the fastest fluctuations.  Time series analysis failed to yield reasonable periodicities and noise analysis did not confirm the presence of Gaussian distributed noise, so we have developed a model based on individual rapid synchrotron pulses caused by a shock wave encountering individual turbulent cells in the jet flow. The size of the vortex region and the magnetic filled orientation of the turbulent cells is stochastic, and results in pulses.  We can decompose the microvariability curve into individual pulses and can get a picture of the underlying turbulent structure.  We were able to fit the 72-hour nearly continuous flux curve in terms of this model and obtain an estimate of the range of sizes of the turbulent cells and the density enhancements.

\begin{acknowledgements}
I would like to thank Florida International University for the financial support through Dissertation Year Fellowship 2012. The Abastumani Observatory team acknowledges financial support by the Georgian National Science Foundation through grant GNSF/ST09/521 4-320. Shao Ming Hu would like to thank the support by the National Natural Science Foundation of China under grants 11143012, 10778619 and 10778701. Rumen Bachev acknowledges the support by Science Research Fund of the Bulgarian Ministry of Education and Sciences through grants DO 20-85 and Bln 13-09.
\end{acknowledgements}

\end{document}